\documentstyle[preprint,aps]{revtex}
\input{psfig}

\begin{document}

\title{Coarse-grained description of thermo-capillary flow}
\author{David Jasnow}
\address{Department of Physics and Astronomy, University of Pittsburgh,
Pittsburgh, Pennsylvania, 15260}
\author{Jorge Vi\~nals}
\address{Supercomputer Computations Research Institute,
Florida State University,
Tallahassee, Florida 32306-4052, and Department of Chemical Engineering,
FAMU/FSU College of Engineering, Tallahassee, Florida 32310}

\date{\today}

\maketitle

\begin{abstract}
A mesoscopic or coarse-grained approach is presented to study
thermo-capillary induced flows. An order parameter representation
of a two-phase binary fluid is used in which the 
interfacial region separating the phases
naturally occupies a transition zone of small width.
The order parameter satisfies the 
Cahn-Hilliard equation with advective transport. 
A modified Navier-Stokes equation that incorporates an explicit coupling to 
the order parameter field governs fluid flow. It reduces, in the 
limit of an infinitely thin interface, to the
Navier-Stokes equation within the bulk phases and to two interfacial
forces: a normal capillary force
proportional to the surface tension and the mean
curvature of the surface, and a tangential force proportional
to the tangential derivative of the surface tension. The method
is illustrated in two cases: thermo-capillary migration of drops
and phase separation via spinodal decomposition,
both in an externally imposed temperature gradient.
\end{abstract}


% Uncomment for twocolumn
%\narrowtext
%]

\section{Introduction}

The study of multi-phase flows leads to a classical moving boundary
problem in which the equations governing fluid motion are solved in each
phase, subject to boundary conditions specified on the moving boundaries.
In the classical case, the boundary of separation is assumed to be an
idealized boundary without any structure. For viscous flows, the
velocity field is continuous across the
boundary, whereas the normal component of the stress tensor is
discontinuous if capillary forces are allowed for.
Implicit in this formulation are, of course,  assumptions
that local thermodynamic equilibrium obtains
and that the typical length scales of spatial structures and
flow are much larger than the scale of the
physical boundary separating the phases. In most cases of interest
to fluid mechanics the scale of the flow is set externally by
macroscopic stresses on the system, whereas the length scale of the
interface is of microscopic size and set by the range of the
intermolecular forces in the fluid. Under these conditions, the
transition region between the two phases can be approximated for all
practical purposes by an ideal, mathematical surface of
discontinuity.

A notable exception to this rule concerns flow in a fluid near or at its
critical point. There the surface of separation between both phases
becomes arbitrarily diffuse, and the flow within the boundary itself
needs to be explicitly
modeled and resolved. The so-called Model~H (following
the nomenclature of Hohenberg and Halperin \cite{re:hohenberg77}) 
was introduced to study
order parameter and momentum density fluctuations near the critical
point, as well as their interaction, at a coarse-grained or mesoscopic
scale \cite{re:siggia76}. 
Additional (reversible or Hamiltonian) terms were added on
phenomenological grounds to the Navier-Stokes equation and to the
equation governing the relaxation of fluctuations in the order parameter
appropriate for the phase transition under consideration. Later studies
based on the same model have
addressed critical fluctuations of simple fluids
\cite{re:onuki79,re:onuki81} and polymers \cite{re:helfand89,re:doi90}
under external shear, and more recently (and more closely
related to our study) hydrodynamic effects on spinodal decomposition in
a binary fluid \cite{re:kawasaki83,re:koga91}.
In spinodal decomposition, a binary system is quenched from a homogeneous
disordered state to a region of the phase diagram in which the homogeneous 
state is linearly unstable to fluctuations in a wide band of length
scales. (For a review, see, e.g., \cite{re:gunton83}.)
Some further details on the general principles behind a mesoscopic description 
of equilibrium properties of bulk phases and of interfacial structures can 
be found in readily available texts and monographs (see, e.g., 
Refs~\cite{re:landau80,re:stanley71,re:jasnow84,re:jasnow86,re:rowlinson82}.)
Implications of such a level of description on the equations
governing fluid flow are discussed in the appendix, where the modifications
of the non-dissipative part of the stress tensor embodied in Model H
are obtained. 
This ``Hamiltonian'' or reactive part of the stress
tensor does not contribute to the entropy increase.
Additional treatments based on
irreversible thermodynamics and continuum mechanics can be found in 
Refs.~\cite{re:gurtin94,re:antanovskii95}.

The usefulness of a coarse-grained approach
away from a critical point is less well established. Previous work along
these lines includes Ginzburg-Landau models used in the study of the 
kinetics of
phase separation (e.g., the Cahn-Hilliard equation) \cite{re:gunton83},
or phase field
models of solidification \cite{re:langer86,re:caginalp85,re:caginalp86}. 
In both cases, a mathematical surface of
separation between the two phases becomes a transition region of
small but finite width, across which all magnitudes change continuously.
The essential ingredient in both cases is the assumption that the
relevant thermodynamic potential density (entropy in 
closed systems, Helmholtz
free energy if the system is held at constant temperature) is a function
not only of the usual thermodynamic variables and of the order
parameter, but also of the spatial gradient of the latter. In the case
of a binary fluid, for example, the entropy density is assumed to depend
on both the mass density of solute and its spatial derivative. As a
consequence, the chemical potential of the solute is no longer given by
the derivative of the local entropy with respect to solute density.

From a theoretical point of view,
it is plausible to assume that modes describing the 
motion of boundaries separating bulk phases,
even away from a critical point, are the long-lived modes
of the dynamics of the system, and hence gradients of the order
parameter can be considered as
additional arguments upon which the thermodynamic potential depends, at
least on the time scale over which the motion of the boundaries is
observable macroscopically. Furthermore, one would also like to satisfy
the consistency requirement that the coarse-grained model reduces to
the classical macroscopic description in the limit of
a small transition region (compared to the scale of relevant structures).
This requires, in general,
relating macroscopically measurable quantities (e.g., the
surface tension of the boundary) to phenomenological
parameters that enter the mesoscopic model \cite{re:fo1}.

One of the chief advantages of a
mesoscopic approach is that it allows the study of fluid
phenomena that lie outside the classical continuum formulation, for
example, the break up and coalescence of fluid domains. More generally, by 
incorporating explicitly into the
model dissipation at short length scales (on the order of the
interfacial thickness or order parameter correlation length), 
the model allows the study of other physical 
situations that may be influenced by phenomena at that scale.
Other examples that may fall in this class include motion
of contact lines, flow near solid walls and slip at a microscopic
or mesoscopic scale.
 
From a purely computational point of view the model used can be
viewed as an extension of a class of numerical methods often used
to study the classical moving boundary problem, namely,
those that phenomenologically 
introduce additional dissipation at short length scales. 
At least in principle, however,  the method that we describe 
in this paper differs
from those approaches (e.g., methods based on the introduction of ad-hoc
\lq\lq artificial dissipation" or \lq\lq artificial viscosity"), since,
in the present case, short length scale dissipation is
intrinsically part of the model, and is related to the order parameter
variable that describes the phases. Also from a computational point of
view, the approach that we follow is similar to the so-called immersed
boundary or diffuse interface methods introduced for the study of multi-phase
flows \cite{re:peskin77,re:hyman84,re:hirt81,re:unverdi92,re:shi94}. 
The order parameter which we define below plays a role analogous
to the ``color function.'' However, it is not an auxiliary field introduced for
computational convenience, but rather a thermodynamic variable in its
own right, controlled by a local free energy density.
As a consequence, for example, the surface tension of this
model cannot be fixed independently, but is completely determined by the
choice of thermodynamic potential and the parameters contained therein.
This will reveal some interesting features to be expected at high
thermal gradients.

In Section \ref{sec:model} we introduce the coarse-grained model used in
our study. We concentrate in this work on thermo-capillary flows and
hence explicitly discuss how to couple the equation governing fluid
motion and relaxation of the order parameter. In this paper we will
consider the case of a constant, externally imposed temperature
gradient, although it is straightforward to extend our calculations to
incorporate a fluctuating temperature field. We also discuss in this
section the limit of a thin interface and the boundary conditions that
result from our model in that limit.
As illustrations, we apply the numerical approach to two problems of 
interest. In Section \ref{sec:bubble} thermo-capillary motion of
small drops under large thermal gradients is considered,
while in Section \ref{sec:spinodal} we study phase separation 
via spinodal decomposition in the presence of a uniform temperature
gradient.  Section \ref{sec:conc} is reserved for concluding remarks.
An appendix provides additional details on the derivation of
modifications to the Navier-Stokes equation introduced through the
coarse-grained level of description.

\section{Coarse-grained model}
\label{sec:model}

\subsection{Formulation}

Consider an incompressible and Newtonian binary fluid at a temperature below
its phase separation 
critical point. Let $\varphi$ be the order parameter density
appropriate for the un-mixing transition, chosen to be $\varphi = 0$
in the disordered phase above the
transition point, and symmetric and equal to $\pm \varphi_{eq}$ below.
[$\varphi$ can be thought of as $c-c_c$, where $c$ is the mass fraction of
one of the species and $c_c$ its value at the phase separation critical
point.] One
further makes the reasonable assumptions for binary liquid systems 
that the shear viscosities of both phases are equal,
and that the dependence of the mass density on the order parameter is weak 
and can be neglected. In effect, we consider here a \lq\lq symmetric" model in 
which all bulk properties of both phases are equal. 
These restrictions, which conveniently dramatize interfacial phenomena in the
cases which we study here, can be easily removed.

If $\vec{v}$ is
the velocity of an element of fluid, conservation of mass and momentum lead
to,
\begin{equation}
\label{eq:incomp}
\nabla \cdot \vec{v} = 0,
\end{equation}
\begin{equation}
\label{eq:ns}
\rho \frac{d \vec{v}}{dt} = - \nabla p +  \eta \nabla^{2} \vec{v} +
\mu \nabla \varphi ,
\end{equation}
where $\mu = \delta {\cal F}/\delta \varphi$ is the chemical potential
conjugate to the order parameter $\varphi$ ($\delta / \delta \varphi$
stands for functional or variational differentiation),
$\rho$ and $\eta$ are the density and shear viscosity of the fluid
respectively, and $p$ is the pressure field.
The last term in Eq.~(\ref{eq:ns}) is non-standard 
in macroscopic approaches and incorporates capillary 
effects as discussed further below and in the appendix. 
As a reference, we note that the Navier-Stokes equation is also modified
in Immersed Boundary Methods by adding a term of the form,
\begin{equation}
\left( \sigma \kappa \hat{n} + \frac{\partial \sigma}{\partial s}
\hat{s} \right) \delta \left( \vec{x} - \vec{x}_{S} \right),
\end{equation}
where $\vec{x}_{S}$ is the instantaneous location of the boundary.
Here $\sigma$ is the interfacial tension (independently prescribed),
$\hat{n}$ and $\hat{s}$ are the unit vectors along the normal and
tangential directions respectively, and $\kappa$ is the mean curvature
of the interface. As we further discuss in Section \ref{sec:sil}, the
term $\mu \nabla \varphi$ in Eq. (\ref{eq:ns}) in the limit of a thin
interface reduces to the term used in Immersed Boundary Methods, but
with $\sigma$ now determined by the coarse-grained free energy
functional ${\cal F}$  (see, e.g.,
Refs~\cite{re:landau80,re:stanley71,re:jasnow84,re:jasnow86,re:rowlinson82})
that, for simplicity, is chosen to be of the form,
\begin{equation}
\label{eq:fe}
{\cal F}\left[ \varphi \right] = \int dV \left\{ \frac{K}{2} | \nabla \varphi
|^{2} + f( \varphi) \right\},
\end{equation}
with,
\begin{equation}
\label{eq:feh}
f(\varphi ) = - \frac{r}{2} \varphi^{2} + 
\frac{\lambda}{4} \varphi^{4}.
\end{equation}
The integration extends over the entire system (both bulk phases and
interfacial regions), and $K, r,$ and $\lambda$ are three phenomenological 
coefficients as yet unspecified, other than requiring $K,\lambda > 0$. 
For the model defined by Eqs.~(\ref{eq:fe}) and (\ref{eq:feh})
the chemical potential is given by 
$\mu = -K \nabla^2 \varphi - r \varphi + \lambda \varphi^3$. 
As is well known, this free energy qualitatively describes a single 
homogeneous phase over a range of parameters ($r < 0$) yielding to 
two-phase coexistence in another region ($r > 0$). 
The parameter $r$ is therefore a measure of distance below the phase separation
critical temperature and is taken to be proportional to that
temperature difference, while the parameter $\lambda$,
at this level of approximation can be related to the inter-particle
potentials (or to a virial coefficient). 
As noted below, the parameters $\lambda$ and $K$ can be determined from
equilibrium measurements. The order parameter is further assumed to
satisfy a modified Cahn-Hilliard equation,
\begin{equation}
\label{eq:ch}
\frac{\partial \varphi}{\partial t} + \nabla \cdot \left( \varphi \vec{v} 
\right) = M \nabla^{2} \mu
\end{equation}
to allow for advective transport of $\varphi$. 
$M$ is a phenomenological
mobility coefficient of microscopic origin, which, in a real system,
could be inferred from mutual diffusion and order-parameter
susceptibility measurements away from the critical region.

Equations (\ref{eq:incomp})-(\ref{eq:ns}) and
(\ref{eq:fe})-(\ref{eq:ch}) completely
specify the model. We have imposed the following boundary conditions at
the edges of the fluid domain: $\vec{v}=0$, $\varphi = \varphi_{eq}(T)$,
and $\partial \varphi / \partial n = \partial \varphi_{eq}(T) / \partial
n$, i.e., the value of the order parameter near the boundary is
determined by the imposed temperature field there. Note
that with this choice of boundary conditions, $\int dV \varphi$ is not a
strictly
conserved quantity. The boundary conditions required to
enforce strict conservation of order parameter are $\partial \varphi /
\partial n = 0$ and $\partial \nabla^{2} \varphi / \partial n = 0$. In
this case small boundary layers of $\varphi$ (of thickness of the order
of the correlation length
$\xi$) would develop because $\varphi$ in the bulk (and away from
interfaces) is approximately equal to $\varphi_{eq}(T)$, and therefore
would not satisfy this latter set of boundary conditions. Our choice of
boundary conditions is motivated by computational simplicity so that we
do not have to resolve additional boundary layers at the edges of the
domains. In any event, the spatial integral of $\varphi$ changes very
little during the course of the numerical solution, and hence this
choice has little effect on the dynamics in the bulk phases.

From Eq.~(\ref{eq:feh}) one finds $\varphi_{eq} = \sqrt{r/\lambda}$
with $\pm \varphi_{eq}$ as the two 
coexisting solutions. 
Equation (\ref{eq:ch}) also admits a nonuniform, \lq\lq kink"
stationary solution for $\vec{v} = 0$ and constant $r > 0$,
\begin{equation}
\label{eq:kink}
\varphi_{0} (u) = \varphi_{eq} \tanh \frac{u}{\sqrt{2} \xi},
\end{equation}
with boundary conditions $\varphi \rightarrow \pm \varphi_{eq}$ as 
$u \rightarrow \pm \infty$. The coordinate $u$ is along an
arbitrary direction and $\xi$ is the mean field correlation length,
$\xi = \sqrt{K/r}$. This stationary solution represents the coexistence
of two equilibrium phases with $\varphi = \pm \varphi_{eq}$, across
a transition region located at $u = 0$ and of width $\xi$. In what
follows, we will refer to the regions $|u /\xi| \gg 1$ as the bulk phases,
and $|u /\xi| \sim 1$
as the interfacial or transition region. In the region
of parameters of Eqs. (\ref{eq:fe}) and (\ref{eq:feh})
that we explore, the surface area
occupied by the bulk phases is much larger than that of the 
interfaces.

The excess free energy density associated with 
Eq.~(\ref{eq:kink}) compared
to the free energy of either uniform phase is given by
\cite{re:rowlinson82},
\begin{equation}
\sigma = K \int_{- \infty}^{\infty} du \left( \frac{d \varphi_{0}}
{du} \right)^{2} = \frac{2K \varphi_{eq}^{2}}{3 \xi}.
\label{eq:surfacetension}
\end{equation}

We introduce dimensionless variables by choosing $\xi$ as the scale of
length, $K/Mr^{2}$ as the scale of time, and $\varphi_{eq}$ as the order
parameter scale. In dimensionless units, the model equations reduce to,
\begin{equation}
\nabla \cdot \vec{v} = 0,
\end{equation}
\begin{equation}
\label{eq:nsd}
{\rm Re} \frac{d \vec{v}}{dt} = - \nabla p + \nabla^{2} \vec{v} + 
{\rm C} \left(  -\nabla^{2} \varphi - \varphi + \varphi^{3}
 \right) \nabla \varphi,
\end{equation}
and,
\begin{equation}
\label{eq:chd}
\frac{\partial \varphi}{\partial t} + \nabla \cdot (\varphi \vec{v}) =
\nabla^{2} \left( -\nabla^{2} \varphi - \varphi + \varphi^{3}
\right).
\end{equation}
Two dimensionless groups remain; one of them,
Re$=Mr/\nu$ ($\nu = \eta / \rho$ is 
the kinematic viscosity) gives the ratio between order parameter
diffusion and momentum diffusion due to viscosity.
In the calculations that we describe in this paper,  the system
is in the overdamped limit of Re$\simeq 0$.
The other, C$= 3 \sigma \xi/2 \eta M r$, plays the role of a capillary number. 
Flows in our study are entirely driven by surface tension. In the 
overdamped limit the velocity scale of such flows is
$v \sim \sigma / \eta$ 
where we have
assumed that the scale of the flow is the same as the scale of the
typical domains of the two phases. Therefore, the characteristic
time for an element of fluid to be advected a distance equal to
the interfacial thickness is $\tau_{\sigma} = \xi \eta / \sigma$. The 
diffusive time scale given above is $\tau_{\varphi} = \xi^{2}/Mr$;
hence C $\propto \tau_{\varphi}/\tau_{\sigma}$.

In order to couple the previous equations to a slowly varying
temperature field, we make, as noted above, the traditional identification 
of the parameter $r$ in Eq.~(\ref{eq:fe})
as being a linear function of temperature
$r(\vec{x}) \propto T_0 - T(\vec{x})$, where
$T_0$ is a constant playing the role of the consolute or critical 
temperature.
In the case that
we study, the temperature field is fixed and remains constant
throughout the calculation. If one wishes to incorporate a fluctuating
temperature field, the model equations have to be supplemented with
the equation of energy conservation \cite{re:fo2}.
In dimensionless units, we consider,
\begin{equation}
\frac{\delta {\cal F}}{\delta \varphi} = - \nabla^{2} \varphi -
 \tau(y)  \varphi + \varphi^{3},
\end{equation}
where the dimensionless local temperature variable $\tau(y) =
\tau_0 + \alpha y$, with 
$\alpha$ the dimensionless temperature gradient.  We identify
the $y-$direction as the direction in which the imposed temperature
varies. For our coarse grained analysis to make sense we require
that temperature variations are small over small distances (of the order
of the interfacial thickness), namely, in dimensionless form, $|\alpha | \ll 
1$, but perhaps large over distances of the order of the system size,
and possibly even over a {\em domain} of one of the phases.

\subsection{Sharp interface limit}
\label{sec:sil}

We next present a heuristic analysis to illustrate that the
coarse-grained model introduced reduces to the classical continuum
description in the limit in which the width of the interfacial region is
much smaller that the size of the bulk domains. We introduce a local
orthogonal system $(s,u)$ such that a given point in space is given by
$\vec{r} = \vec{R}(s) + u \hat{n}$, where $\vec{R}(s)$ describes the
location of the level set $\varphi(\vec{r})=0$ ($s$ is the arclength
along it), $\hat{n}$ is the local normal pointing towards the (+) phase,
and $u$ is the distance along the normal direction.

In the bulk regions ($|u/\xi | \gg 1$) $\mu \nabla \varphi$ is of order 
higher than linear in the gradients and therefore negligible. One then
recovers the standard Navier-Stokes equation.
The Cahn-Hilliard equation can be likewise
linearized around $\varphi_{eq}$ in the bulk regions, yielding the
standard diffusion equation with advective transport.

In the interfacial regions ($|u/\xi | \sim 1$), the term 
$\mu \nabla \varphi$ does become large. In the limit of gently curved 
interfaces, $| \kappa \xi|  \ll 1$, where $\kappa$ is the 
mean curvature of the interfacial region,
and for small temperature gradients, $ | \alpha | \ll 1$, the field
$\varphi$ relaxes on a fast time scale to essentially the stationary
solution $\varphi_{0}(u)$ as in Eq.~(\ref{eq:kink}) ~\cite{re:boyanovsky95}.
One can develop a multiscale expansion of the field $\varphi
= \varphi(S,u,U)$ with $S,U$ the slowly varying coordinates in the
tangential and normal direction as set by the slow variation of the
temperature field, and $u$ the fast variation in the normal direction as
determined by the free energy functional. We further assume -- and this
remains to be proven more rigorously -- that
$\varphi$ relaxes quickly to its local equilibrium profile given by the
local value of the temperature, but there remain slow gradients of
$\varphi$, since this approximate solution is no longer a solution of
$\delta {\cal F}/\delta \varphi =~\mbox{constant},$ 
throughout the system.
The surface tension in Eq.~(\ref{eq:surfacetension}) can then be written as,
\begin{equation}
\sigma = \int_{-\infty}^{+\infty} du \Delta F(u,U,S),
\end{equation}
where $F = \frac{1}{2}K|\nabla \varphi|^2 + f(\varphi)$ and
$\Delta F$ is the free energy (density) difference between the mean field
value of $F$ at the local value of $\varphi$ in
a configuration with a two-phase interface and the free energy of
the bulk phases $\varphi_{eq}$, at the local temperature. Under the
assumptions stated above, this expression can be used to define a slowly
varying surface tension $\sigma(S)$ along the interface, since the 
magnitude of the order parameter
$\varphi$ will be the local equilibrium value at the local temperature.
Then,
\begin{equation}
\frac{\partial \sigma}{\partial S} = 
\int_{-\infty}^{\infty} du
\left( \frac{\partial F_{int}}{\partial \varphi} \right) \frac{\partial
\varphi}{\partial S} = \int_{-\infty}^{\infty} du \mu \frac{\partial
\varphi}{\partial S} \, \, ,
\label{eq:tang_comp}
\end{equation}
where we have used the fact that the metric tensor element $g_{22} =
\frac{d\vec{r}}{ds} \cdot \frac{d\vec{r}}{ds} = (1 + u \kappa )^{2}
\approx 1$ in the limit of gently curved interfaces 
considered. (Note that $\kappa < 0 $ for a sphere). The chemical potential
appearing is appropriate for an interfacial configuration and is
non-vanishing for a gently curved
(or indeed a flat) interface in a temperature gradient.
Hence the tangential component of
$\mu \nabla \varphi$ gives rise to a tangential surface force which
equals the tangential derivative of the surface tension.

We can likewise study the normal component by recalling that,
\begin{eqnarray}
\int_{-\infty}^{\infty} du \mu \frac{\partial \varphi}{\partial u} & = &
\int_{-\infty}^{\infty} du \left[ - \nabla^{2} \varphi + f' \right]
\frac{\partial \varphi}{\partial u} \nonumber \\ 
{} & \simeq & \int_{-\infty}^{\infty} du
\left[ - \frac{\partial^{2} \varphi}{\partial u^{2}} - \kappa(S)
\frac{\partial \varphi}{\partial u} - \frac{\partial^{2}
\varphi}{\partial S^{2}} + f' \right] \frac{\partial \varphi}{\partial
u} \nonumber \\
{} & \simeq & \kappa (S) \int_{-\infty}^{\infty} du \left( \frac{\partial
\varphi}{\partial u} \right)^{2} - \int_{-\infty}^{\infty} du
\frac{\partial^{2} \varphi}{\partial S^{2}} \frac{\partial
\varphi}{\partial u}.
\label{eq:normal_comp}
\end{eqnarray}
The first term gives again the surface tension $\sigma$, and the second
term would remain in the sharp interface limit due to the slow variation
of $\varphi$ with temperature.

Therefore in the case of a thin interfacial region, the additional term
in the Navier-Stokes equation is equivalent -- under the stated
assumptions of a slowly varying temperature field and gently curved
interfaces -- to a tangential surface force proportional to the tangential
derivative of the surface tension, and to a normal force proportional to
the mean curvature of the interface.

\subsection{Numerical method}

We next discuss the numerical algorithm used in the actual computations.
We restrict our attention in this paper to two dimensional flows. Since
the velocity field is solenoidal, it is convenient to introduce the
stream function,
\begin{equation}
\vec{v} = \nabla \times (\psi \hat{k}) = \frac{\partial \psi}{\partial y}
\hat{i} - \frac{\partial \psi}{\partial x} \hat{j},
\end{equation}
where the velocity field is defined in the $(x,y)$ plane, and $\hat{k}$
is the unit normal perpendicular to that plane. By taking the curl of
Eq. (\ref{eq:nsd}) with Re = 0 we find,
\begin{equation}
\label{eq:biharpsi}
\nabla^{4} \psi + {\rm C} \left[ \nabla \left( \nabla^{2} \varphi \right)
\times \nabla \varphi \right] \cdot \hat{k} = 0.
\end{equation}
The flow field can be found by solving a biharmonic equation for the 
stream function, subject to the boundary conditions that both
the stream function $\psi$ and its normal derivative
$\partial \psi / \partial n$ vanish on the external boundaries of the
system. 

Concerning the equation for the order parameter, Eq.~(\ref{eq:chd}),
we follow Ref.~\cite{re:bjorstad84} and use a backward implicit
method which is first order in time,
\begin{equation}
\label{eq:backward}
\frac{\varphi (t + \Delta t ) - \varphi ( t ) }{\Delta t} +
\nabla \cdot (\vec{v}(t) \varphi (t + \Delta t)) = - {\cal L}
\varphi (t + \Delta t) + \nabla^2 \varphi^{3} (t + \Delta t),
\end{equation}
with  ${\cal L} = \nabla^{4} + \nabla^{2}$. The convective term has been
kept in conservative form, and the velocity field retained in the 
discretization is at time $t$. The boundary conditions that we use for
the order parameter are
$\varphi = \varphi_{eq}(T) $ and $\partial \varphi / \partial n =
\partial \varphi_{eq}(T) / \partial n$ on the external boundaries.
A Gauss-Seidel iteration is used to solve Eq. (\ref{eq:backward}).
First, one considers an \lq\lq outer" iteration,
\begin{equation}
\label{eq:outer}
\varphi (t + \Delta t) \simeq \varphi_{k+1} = \varphi_{k} + \delta_{k} 
\quad {\rm with} \quad \varphi_{0} = \varphi (t).
\end{equation}
Substituting $\varphi (t + \Delta t)$ in Eq. (\ref{eq:backward}) 
and linearizing
in the outer correction field $\delta_{k}(x,y)$ yields an equation for
the outer correction with a residual that is a function of 
$\varphi_{k}$.
Successive iterates converge to a solution when both the residual and
$\| \delta_{k} \|$ go to zero simultaneously. The outer iteration is performed
simultaneously on Eq. (\ref{eq:biharpsi}) and (\ref{eq:backward}).
In practice, an \lq\lq inner" iteration is set up to solve for
$\delta_{k}$, by assuming,
\begin{equation}
\label{eq:inner}
\delta_{k} \simeq \delta_{k,m+1}= \delta_{k,m} + \eta_{m},
\end{equation}
where $\eta_{m}$ is the inner correction. The variable coefficient
biharmonic operator acting on $\eta_{m}$ is then approximated by a 
constant coefficient biharmonic operator,
\begin{equation}
\label{eq:biharphi}
\left[ \nabla^{4} + a \nabla^{2} + b \right] \eta_{m} = R_{inner,m}^{\varphi},
\end{equation}
where $a = 1 - 3< \varphi_{k}^{2} >$ and $b = 1/\Delta t - 6
< \varphi_{k} \nabla^{2} \varphi_{k} >$, and $< {\bf \cdot} >$ stands for
spatial averages over the entire system. Then both Eq. (\ref{eq:biharpsi})
and (\ref{eq:biharphi}) are solved with a fast biharmonic solver 
(see ref. \cite{re:bjorstad84} or \cite{re:chella94} for additional
details).

\section{Thermo-capillary motion of drops}
\label{sec:bubble}

In this section we briefly summarize application of the computational
method introduced in Section~\ref{sec:model} to the thermo-capillary flow of 
a drop of one phase in the 
background of its coexisting partner. As we have noted above, the
main strengths of the coarse grained technique are that (i) it allows
natural tracking of the interface separating the two phases, (ii) it 
naturally encompasses topology changes, such as the merging of 
two droplets, and (iii)
it can include phenomena from the scale of the thermal correlation
length up to macroscopic while providing a qualitatively correct picture
of behavior near criticality.
Accordingly, to separate effects from those
which are normally treated at the macroscopic level, we turn to 
relatively small drops and relatively large temperature gradients.

To illustrate the method we consider the thermo-capillary flow of
a single drop. Initially the radius of the small drop is fixed
at $R \sim 10 \xi,$ where $\xi$ is the thermal correlation length
at our reference temperature $\tau_0 = 1$, which is typically the
value of the temperature parameter at the cold end. This size drop
is about as small as we can go within the coarse grained description
and retain some expectation that the qualitative results will translate
to the macroscopic scale. However on a macroscopic scale,
we consider temperature gradients which are large, so that the 
dimensionless number is $R |\nabla T|/T \sim 0.01$. Hence, over a 
correlation length, the temperature change is typically 
0.1\%, allowing us to retain the coarse grained approach, but over
a drop, the change is on the order of 1\%. By macroscopic standards
this is large. Consider, for example, a nucleating  $1 \mu m$ drop 
in a binary liquid system
in a temperature gradient of 1~C/mm at roughly room temperature.
The dimensionless ratio is three orders of magnitude smaller.

\begin{figure}
\psfig{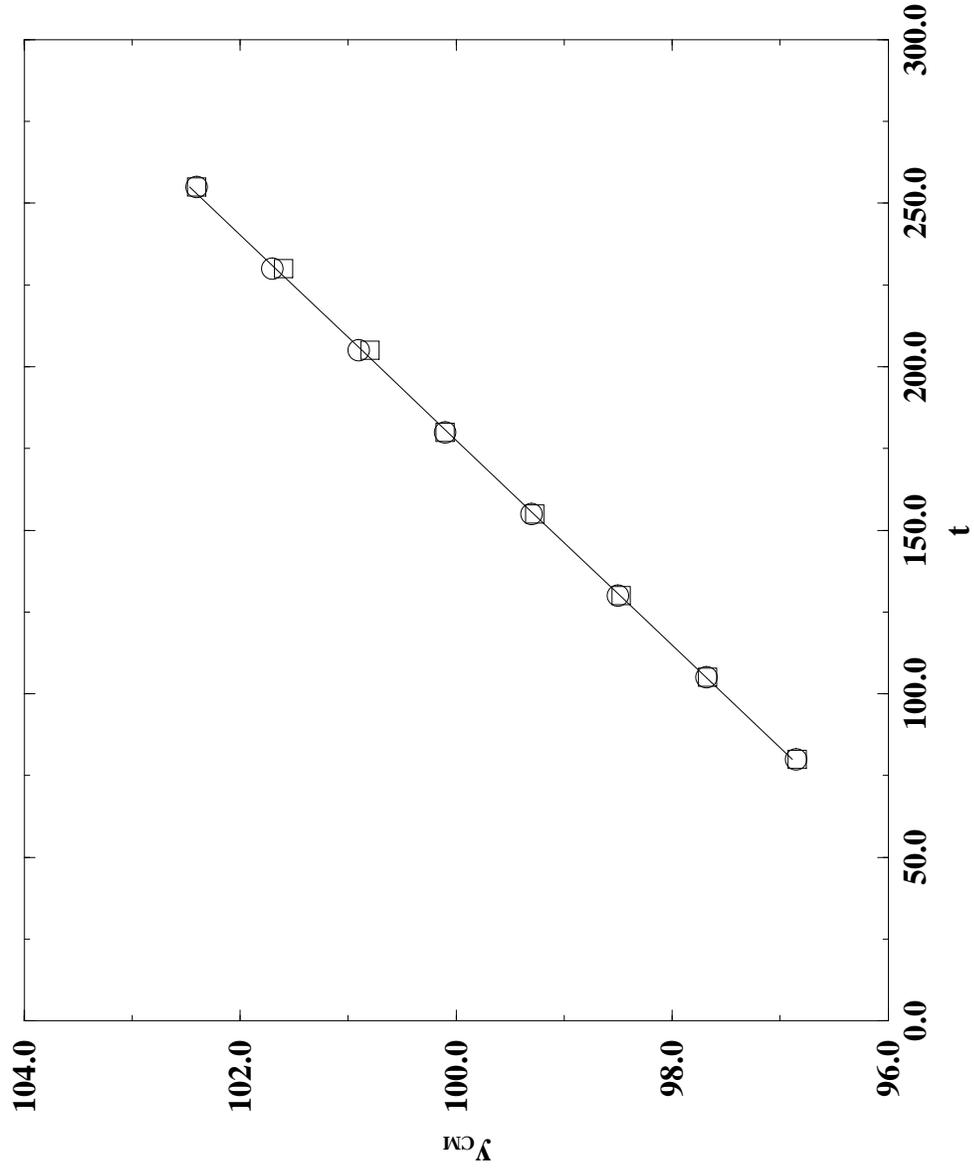}
\caption{Displacement of the drop's center of order parameter as a function
of time for $C = 1, R=12, \tau_{0} = 1$ and $\alpha = -0.004$ in a 
square cell of side $a = 200$. Values for two system sizes are shown:
$\circ , ~ a=400$;  $\Box , ~ a = 200$.
The straight lines are linear fits to the data with slopes
(dimensionless velocities) 0.0319 
($a=400$) and 0.0316 ($a=200$) respectively.}
\label{fi:vlinear}
\end{figure}

\begin{figure}
\psfig{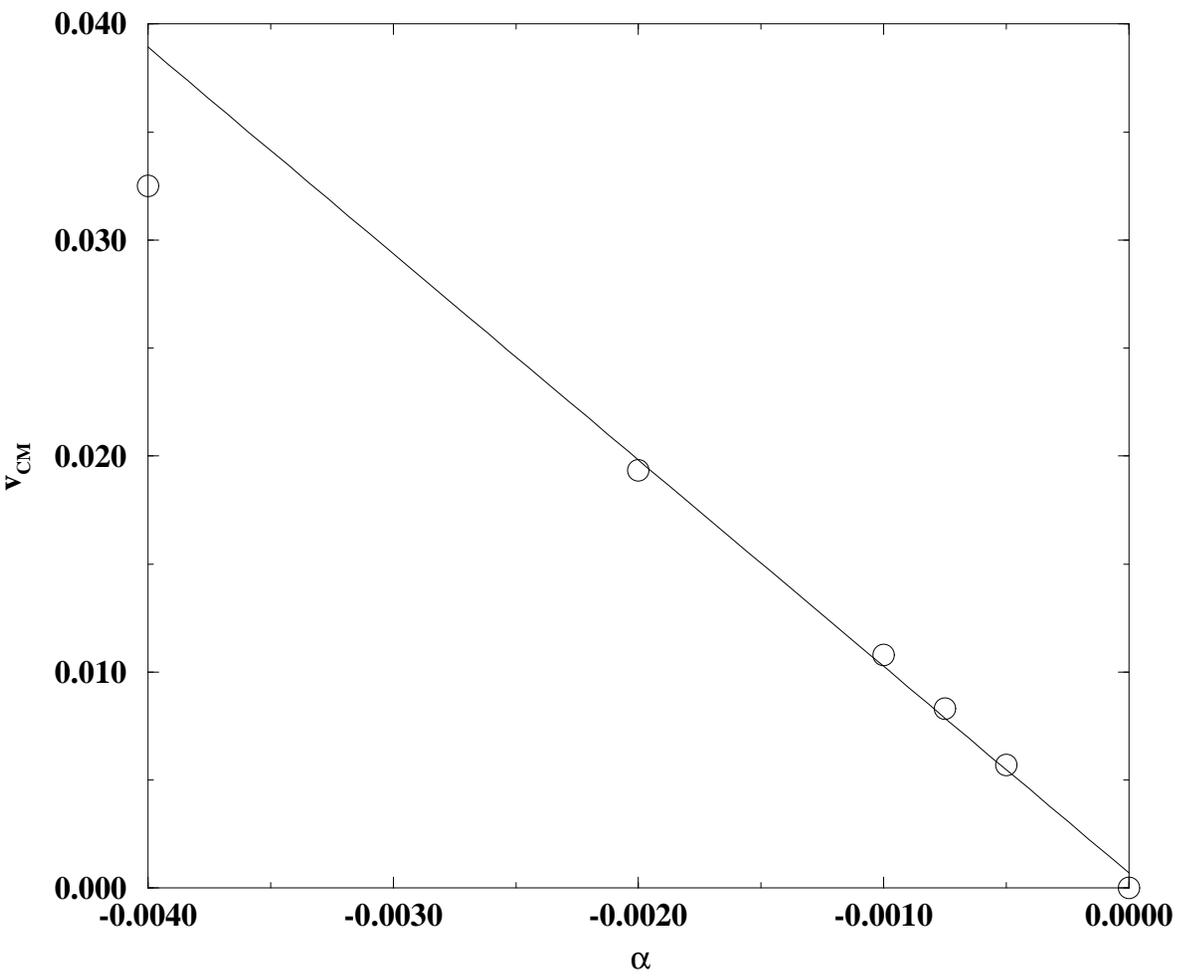}
\caption{Velocity of the center of order parameter versus temperature gradient
for $C = 1, R = 12$ and $\tau_{0} = 1$ in a square cell of side $a=200$.}
\label{fi:vgradt}
\end{figure}

Our first diagnostic is the drift velocity of such a droplet
as a function of the temperature gradient.
In all figures to follow, the high temperature side is at the
bottom. First a sample 
plot of the droplet \lq\lq center of order parameter" (defined by analogy
with the center of mass)  as a function
of time is shown in Fig.~\ref{fi:vlinear}, indicating motion with 
constant velocity. We have repeated the calculation for a number of
values of the temperature gradient $\alpha$, and the results are shown
in Fig.~\ref{fi:vgradt}. At low values of the
gradient, we observe linearity~\cite{re:young59,re:subramanian90} of velocity, 
but at our higher values there is a reproducible reduction as seen in 
the figure.~\cite{re:foot_dim}
Furthermore, at the higher values of the gradient the velocity
of the drop is dependent on the temperature itself,
as well as the gradient. These features
are associated with the fact that this model, which can (qualitatively)
apply near the phase separation critical point as well, has an
equilibrium order parameter depending
on the dimensionless temperature parameter as 
$\tau^{1/2} \sim (T_0 - T)^{1/2}$,
as well as a nonlinear dependence of the interfacial tension,
$\sigma \sim \tau^{3/2} \sim (T_0 - T)^{3/2}$.
These effects, for fixed gradient, become
relatively more apparent near the high temperature side.

\begin{figure}
\psfig{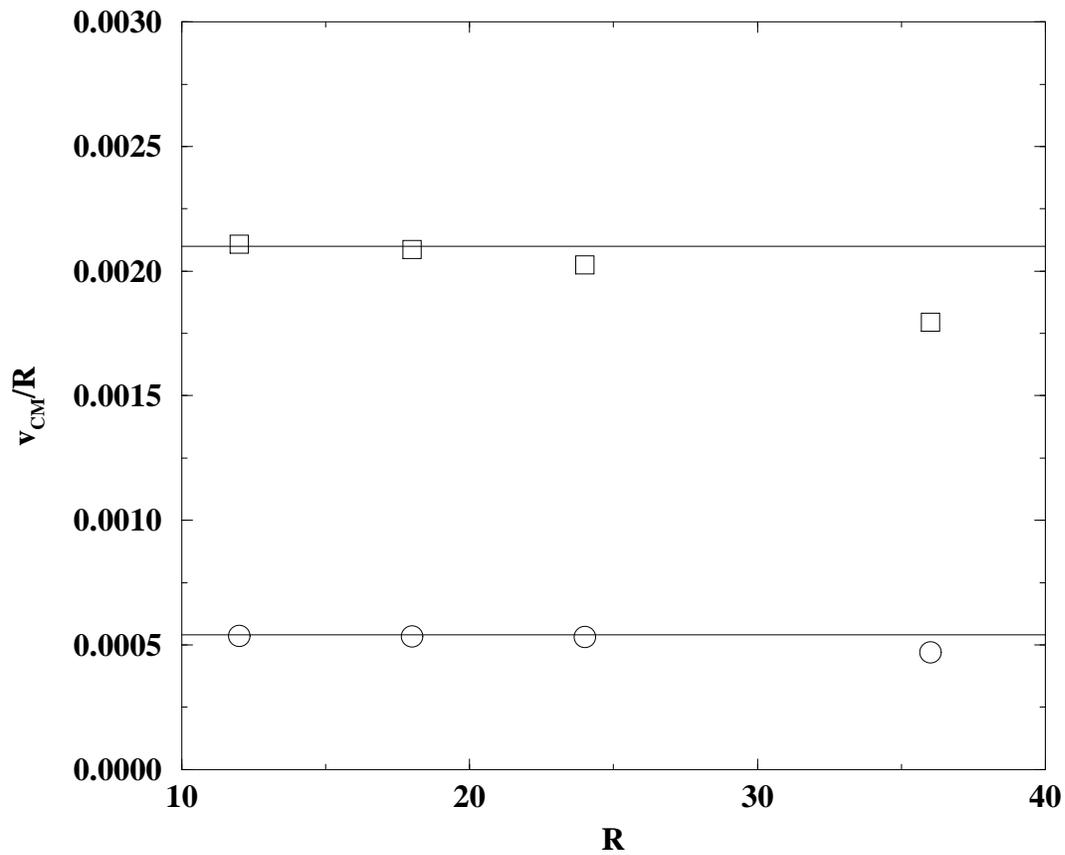}
\caption{Velocity of the center of order parameter as a function of the 
drop radius for  $C=1, \tau_{0} = 0.5$ in a square cell of side  $a=200$:
$ \circ , ~ \alpha = -0.001$;  $\Box , ~ \alpha = -0.004$. Note that $v_{CM}$
is proportional to $R$ for small values of $R$. Interestingly, linearity is lost
at lower values of $R$ at larger temperature gradients.}
\label{fi:vradius}
\end{figure}

Despite the fact that the drop velocity for high gradients reflects
nonlinearities inherent in the model, we have found that, for small
enough temperature gradient, the
velocity scales quite linearly with drop radius over a limited range
(a factor of three)
accessible in these preliminary studies. For larger gradients,
the explicit dependence of both order parameter and surface tension
on temperature
eventually affects the proportionality as seen in Fig.~\ref{fi:vradius}

Even at the relatively high gradients (by macroscopic standards)
used we observe little distortion of the droplets. Discussion
of this issue, further details on the flow fields and analytic analysis
of the droplet migration analysis based on the coarse grained equations
is beyond the present scope.

\begin{figure}
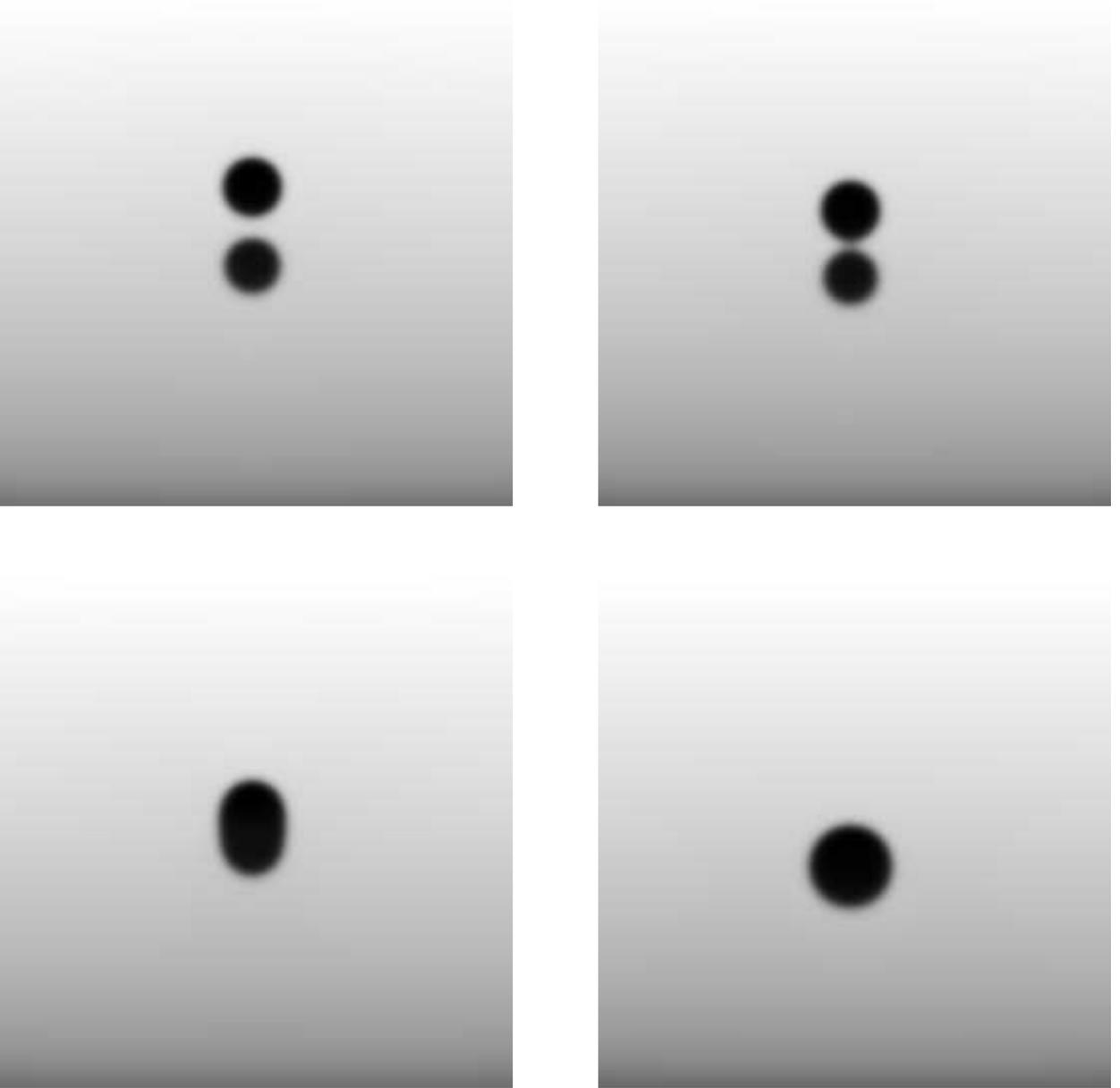

\hbox{
\psfig{figure=dj1.fig5a,width=3in}
\hspace{1cm}
\psfig{figure=dj1.fig5b,width=3in}
}
\vspace{1cm}
\hbox{
\psfig{figure=dj1.fig5c,width=3in}
\hspace{1cm}
\psfig{figure=dj1.fig5d,width=3in}
}
\caption{Sequence of configurations showing the coalescence of
two drops of radius $R = 12$ in a cavity with $a = b = 200$. In this
case $C = 1, \alpha = -0.004$ and $\tau_{0} = 1$. The dimensionless
times shown (from left to right and top to bottom) are $t=1305, 1555,
1605$ and 2055.}
\label{fi:two-bubbles}
\end{figure}

We have finally done a qualitative study of drop coalescence in a
temperature gradient. A sequence of configurations illustrating the motion
of two drops is shown in Fig.~\ref{fi:two-bubbles}. Coalescence involves a 
topology change, which is naturally handled within the coarse-grained approach.
An analysis of the kinetics of droplet coalescence, and droplet 
detachment and attachment to boundaries, is a potentially
rich area but also beyond the present scope. 

\section{Spinodal decomposition in a temperature gradient}
\label{sec:spinodal}

In order to study whether the method can describe more complex
flows, we have studied spinodal decomposition of a binary fluid, in two
spatial dimensions, and under an imposed constant temperature
gradient. 
In dimensionless units, the size of the rectangular
fluid domain studied is
$a = 800$ and $b = 200$ (in the gradient ($y$) direction).
We have used an evenly spaced grid
with $m=1024$ nodes along the $x$ direction, and $n=256$ along
the $y$ direction. The results presented involve a dimensionless
temperature gradient $\alpha = -0.004$ 
along the $y$ direction, and
include up to $t = 100$. The initial condition for the
order parameter at each point is a set of uniformly
distributed random numbers in $[ -0.01, 0.01 ]$. The time step
used in the numerical integration is $dt = 0.25$.

\begin{figure}
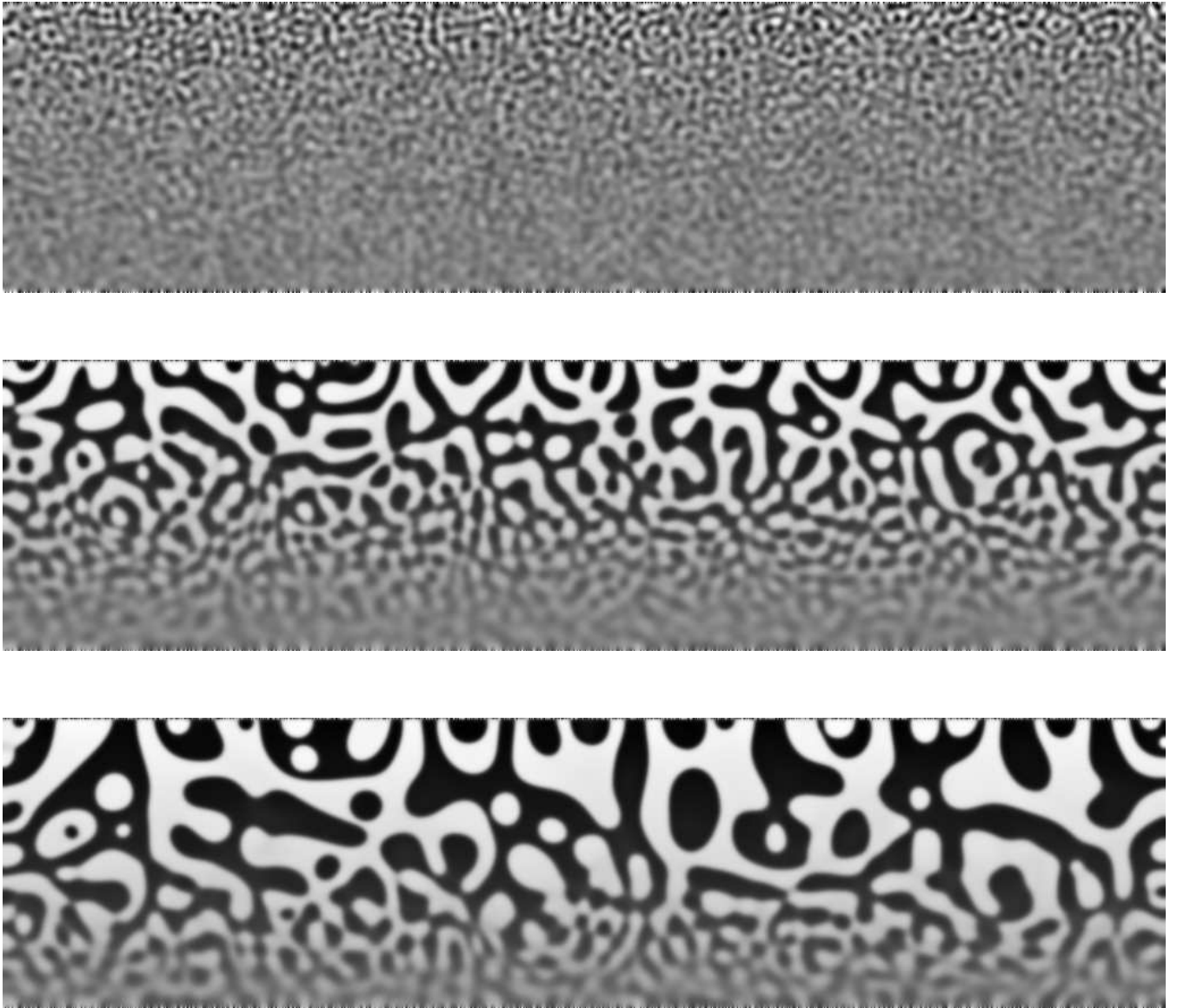

\psfig{figure=dj1.fig6a,width=7in}
\vspace{1cm}
\psfig{figure=dj1.fig6b,width=7in}
\vspace{1cm}
\psfig{figure=dj1.fig6c,width=7in}
\caption{Spinodal decomposition of a binary fluid in a rectangular cell of
dimensions $a = 800$ and $b = 200$ (vertical direction in the figures, 
corresponding to the $y$-direction in the fluid). The parameters used are
$C = 1, \tau_{0} = 1$ and $ \alpha = -0.004$. Note that the hotter end is at
the bottom of each figure. Shown is the value of the order parameter in 
grey scale at three instants of time (from top to bottom): $t = 6.25, 50$ 
and 100.}
\label{fi:configurations}
\end{figure}

Figure \ref{fi:configurations} shows an example of evolution of
the system for three different instants of time: $t = 6.35, 50$ and 100.
The order parameter (between -1 and 1 in the dimensionless units we are 
using) is shown
in grey scale. The characteristic spinodal pattern emerges, with
an intensity and domain size that is a function of the local 
temperature of the 
system. The temperature parameter $r$ as introduced in Eq.~(\ref{eq:fe})
is known to couple extremely weakly to the phase separation
process. It is not expected to change the algebraic
form of the growth law for
the domains (i.e., the power-law growth in time of the 
characteristic domain size), but rather it may change the amplitude.
Two questions naturally arise for a 
slowly varying temperature, namely, whether the 
temperature gradient introduces any anisotropy in the characteristic
length scale of the pattern, and whether an
effective growth law {\em amplitude} exists that slowly changes in 
space, as the temperature of the system changes. Roughly speaking,
the question is whether in the presence of a
slowly varying temperature, phase separation proceeds temperature
\lq\lq slice" by \lq\lq slice" as in ordinary spinodal decomposition at that 
final temperature. Directional anisotropy appears to arise in the purely
diffusive Model B (conserved order parameter
without hydrodynamic coupling, \cite{re:hohenberg77}) 
at early to intermediate times
(in the form of a lag in the growth parallel to the gradient),
and it carries over into the late stage growth stage.~\cite{re:llambias94}
Preliminary results over a small ensemble of independent 
initial conditions (five) indicate that, contrary to
the situation in Model B, there is no appreciable 
anisotropy in the characteristic length scale of the domains. It is 
interesting to note that even if thermo-migration effects themselves
are small for dimensionless times $t \leq 100$, the hydrodynamic coupling
and complex flows
appear to wash out any prominent anisotropy.   

\section{Conclusions}
\label{sec:conc}

We have implemented a numerical algorithm to solve the coupled
Cahn-Hilliard and extended Navier-Stokes equations (Model~H) in a
temperature gradient. The backward implicit method is unconditionally
stable, and with moderate computing effort has allowed us to study
reasonably large system sizes for a reasonable amount of time. Two
particular examples of thermo-capillary induced flow
in two dimensions have been studied:
the motion of droplets of one phase in the background of its
coexisting partner
in a temperature gradient, and phase separation via spinodal
decomposition, also in a temperature gradient.
In two dimensions a stream function representation has been used
so that an efficient biharmonic solver handles both the order parameter
and flow equations together.
In the first case, the
results given by this method agree with classical macroscopic
calculations in that the migration velocity of the drop is proportional
to both the imposed temperature gradient and the radius. For large
gradients, the velocity depends nonlinearly on the temperature gradient,
as it is to be expected given the dependence of the miscibility gap and
surface tension on temperature in the modeling of the coarse grained
free energy.

The chief advantage of the method is that it allows the study of the
flow on length scales which are not accessible to classical macroscopic 
methods. One such case involves the coalescence of drops and the 
concomitant topology change of the interfaces. We have presented
qualitative results involving the coalescence of two drops induced by
thermo-capillary migration. 
Of course, the mesoscopic or coarse-grained
model introduced specifies a dependence of the fluid variables at
such length scales, and this requires the introduction of a
phenomenological free energy at that coarse-grained scale.
This coarse-grained free energy is a function of a
few phenomenological parameters (imagined to depend smoothly on the
temperature and other experimentally accessible variables)
and couples through the associated
chemical potential to the velocity field.
The requirement that the model
equations lead to the classical, macroscopic description in the limit of
a small interfacial region allows one, in principle, to determine 
some of the phenomenological parameters (at least within a 
mean field context).

We have also argued that the model introduced reduces to the classical
boundary conditions at the two-phase interface: namely, a normal 
force given by
the surface tension of the interface and its mean curvature, and a
tangential force equal to the tangential derivative of the surface
tension. These boundary conditions follow in the limit in which the
imposed temperature field does not change appreciably over the scale of
the interface. For these arguments
we have used the plausible assumption (consistent with observations from 
simulations)
that the order parameter within the two phases and in the interfacial
region relaxes quickly to its local equilibrium value determined
by the local curvature of the interface and the local temperature. A
more systematic analysis to elucidate this point is clearly needed. This
analysis can also lead to a systematic cataloging of corrections to the
macroscopic boundary conditions.

An additional advantage of the method is that it allows the study (at
least qualitatively) of very small drops and large gradients. In fact,
this method is not likely to be competitive with classical methods in
situations in which gradients are weak and drops are large, as compared to
the scale of the interface. We remark, however, that classical methods
cannot handle topology changes that
are controlled by physical processes on short or
intermediate length scales, such as the order-parameter correlation
length.
Finally, there are several classes of
problems in which the behavior at short lengths scales needs to be
resolved since it eventually determines the flow at macroscopic scales.
Examples include the motion of contact lines, discontinuous velocity
fields at boundaries, and slip and thermal fluctuations at the
mesoscopic scale.

\section*{Acknowledgments} The work of DJ is supported by the Microgravity 
Science and Applications Division of NASA under grant NAG3-1403. The
work of JV is supported by the U.S. Department of Energy, contract No.
DE-FGO5-95ER14566, and also in part by the Supercomputer Computations Research
Institute which is partially funded by the U.S. Department of Energy, 
contract No. DE-FC05-85ER25000. Most of the computation was carried
out using the facilities of the NCCS at the Goddard Space Flight Center.

\appendix

\section{Coarse-grained fluid mechanics}

In this Appendix, a derivation is presented of the modifications of the 
Navier-Stokes equation that are introduced if the description of the
system goes beyond local thermodynamics. In other words, if 
the density or order parameter varies on short length scales so that
a square gradient or similar description is appropriate 
(see, e.g., 
Refs~\cite{re:landau80,re:stanley71,re:jasnow84,re:jasnow86,re:rowlinson82}), 
the Navier-Stokes equation contains new terms in the dissipationless
(reactive or Hamiltonian)
part of the stress tensor. To find these terms we can consider
an ideal fluid. For this derivation we use a Hamiltonian (canonical)
formalism since, we believe, it is simplest. A more detailed
discussion of the Hamiltonian formalism is contained in
the book by Zakharov et al.~\cite{re:zakharov92}, which we draw on;
a discussion of variational principles for an Euler fluid is 
contained in the monograph by Serrin.~\cite{re:serrin59}

Following Zakharov et al. we consider an inviscid fluid with 
\begin{eqnarray}
\partial_{t} \rho + \nabla \cdot (\rho \vec{v}) & = & 0 \nonumber \\
\partial_{t} \chi_i + \vec{v} \cdot \nabla \chi_i & = & 0
\label{eq:conservation}
\end{eqnarray}
where $\vec{v}(\vec{r},t)$ is the Eulerian velocity field, $\rho(\vec{r},t)$
is the mass density and the scalar functions
$\chi_i, \,\,\, i = 1, 2, \dots$, represent
any advected quantities
carried with the flow and satisfying $d\chi/dt=0$. 
The entropy per unit mass satisfies $ds/dt=0$ in
isentropic flow so that we may treat $s=$ constant.  
This allows the pressure to be expressed as $p = p(\rho)$
and the {\em local part} of the
energy per unit volume may be taken 
as $\epsilon(\rho,s) \rightarrow \epsilon(\rho)$.
For a single component fluid the function $\chi$ represents an advected
velocity circulation. Later when we consider a binary fluid, in the
absence of diffusion, the order parameter $\varphi$, representing the 
mass fraction of one species, will be such a quantity. As usual the
local part of the enthalpy per unit mass is 
$h = \partial \epsilon/ \partial\rho$. The energy of the fluid is taken to be
\begin{equation}
{\cal H} = \int [ \frac{1}{2}\rho v^2 + \epsilon(\rho) ] d\vec{r}.
\label{eq:hamiltonian}
\end{equation}
Following Ref.~\cite{re:zakharov92} we show that standard methods reproduce the
Euler equation. Then we modify Eq.~(\ref{eq:hamiltonian}) to go beyond
local thermodynamics to find the new contributions to the stress tensor.

A Lagrangian is defined in a standard fashion introducing Lagrange 
(undetermined) multipliers for the conservation of $\rho$ and $\chi$,
\begin{eqnarray}
{\cal L} & = & \int \left\{ \frac{1}{2} \rho v^2 - \epsilon(\rho) 
 +\Phi \left[ \partial_t \rho + \nabla \cdot (\rho \vec{v}) \right]
\right.  \\
{} & {} & \left. -\lambda \left[ \partial_t \chi + (\vec{v} \cdot \nabla \chi )
\right] \right\} d \vec{r}
\end{eqnarray}
The total action ${\cal S} = \int {\cal L} dt $ is presumed to be an 
extremum with respect to flows $\vec{v}$ which yields
\begin{equation}
\vec{v} = \nabla \Phi + \frac{\lambda}{\rho} \nabla \chi
\label{eq:vel}
\end{equation}
Variation with respect to the multipliers $\Phi$ and $\lambda$ 
recovers the conservation conditions in Eqs.~(\ref{eq:conservation}).
Substitution of the velocity back into the Lagrangian yields the
hamiltonian (canonical) description 
\begin{equation}
{\cal L} = \int \left[ \Phi \partial_t \rho - \lambda \partial_t \chi 
\right] d \vec{r} -{\cal H}
\end{equation}
where ${\cal H}$ is expressed in terms of the conjugate pairs
$\rho, \Phi$ and $\lambda, \chi$. The Hamiltonian equations of motion are
\begin{eqnarray}
\partial_t \rho & = & \delta {\cal H}/\delta \Phi  \nonumber \\
\partial_t \Phi & = & -\delta {\cal H} / \delta \rho \\
\label{eq:ham_eqs}
\partial_t \lambda & = & \delta {\cal H}/\delta \chi \nonumber \\
\partial_t \chi   & = & - \delta {\cal H}/\delta \nonumber \lambda
\end{eqnarray}
         
It is straightforward to show that the second and third equations yield
\begin{eqnarray}
\partial_t \Phi + \vec{v} \cdot \nabla \Phi - \frac{1}{2} v^2 + h = 0
\label{eq:phidot} \\
\partial_t \lambda + \nabla \cdot (\lambda \vec{v}) = 0
\end{eqnarray} 
Applying $\nabla$ to the first yields the familiar Euler equation
for the velocity change
\begin{equation}
\partial_t \vec{v} + \vec{v} \cdot \nabla \vec{v} = -\rho^{-1} \nabla p
\end{equation}
using the local thermodynamics $\partial h/\partial p = \rho^{-1}$.
This is easily generalized to include external potentials yielding
additional body forces.

Now if the physics dictate that one must go beyond local thermodynamics
with, for example, a square gradient, the Hamiltonian~(\ref{eq:hamiltonian})
is modified to read
\begin{equation}
{\cal H} = \int \left[ \frac{1}{2}\rho v^2 + \epsilon(\rho)
+ \frac{1}{2} K(\rho) (\nabla \rho)^2 \right] d \vec{r}.
\label{eq:sq_grad}
\end{equation}
Following all the same steps yields the modified Euler equation
\begin{equation}
\partial_t \vec{v} + \vec{v} \cdot \nabla \vec{v}
 = \rho^{-1} \left[ - \nabla p + \rho \nabla (\nabla \cdot K \nabla \rho )
-\frac{\rho}{2} K'(\rho) (\nabla \rho )^2 \right]
\end{equation}
Terms other than $\nabla p$ can be thought of as additional
body forces owing to short length scale variations of the density.
The dissipationless part of the stress tensor is appropriately
modified. In the most general case, any
functional $g(\rho, \nabla \rho)$ can be added to the
local energy in Eq.~(\ref{eq:sq_grad}), and the stress tensor is 
modified accordingly.  The modification of the stress tensor for
a single-component fluid at the square-gradient level was derived
by Felderhof~\cite{re:felderhof70} using a Lagrangian (particle) 
description of the mechanical equations. 

Further discussion of the generality of variational formulations
of inviscid hydrodynamics are contained in the
Refs.~\cite{re:zakharov92,re:serrin59}.
The addition of one additional advected quantity, representing the
\lq\lq conservation of the identity of fluid particles" will allow 
every flow to be represented as an extremal for the 
(Herivel-Lin) variational principle.~\cite{re:serrin59}

Now we turn to the more interesting case of a two-component fluid. 
Background is contained in the text of Landau and 
Lifschitz~\cite{re:landau59}.
The total momentum flux is $\rho \vec{v}$, where $\rho$ is the total
mass density. The density of one of the species is $\rho c$ where
$c$ is the (dimensionless) mass fraction, which
plays the role of the order parameter (denoted $\varphi$ in 
the body). Now the local energy per
unit volume becomes $\epsilon(\rho, c, s)$, and for isentropic
flow of an ideal fluid we may neglect the specific entropy.
In the absence of diffusion we have
$\partial_t c  + \vec{v} \cdot  \nabla c = 0 $ so that we may consider
$c$ as an advected quantity, denoted generally by $\chi$ above.
For simplicity (and realistically for typical binary liquid
systems) we consider modifications of local thermodynamics 
due to short length-scale concentration variations only. 
In other words, for the present purposes
we assume, as is normally the case, that any density
variations are on scales large compared to the
order-parameter correlation length, $ \sim \xi,$ so that a
square-gradient contribution for the total density is not required.
This assumption may be generalized in special circumstances. 

Hence,
the total energy of the system is taken to be
\begin{equation}
{\cal H} = \int \left[ \frac{1}{2}\rho v^2 + \epsilon(\rho,c)
+ \frac{1}{2} K(c) (\nabla c)^2 \right] d\vec{r}.
\label{eq:sq_grad_c}
\end{equation}
In Hamilton's equations (A7)
the mass fraction $c$ now plays the role of the advected $\chi$ as noted.
Now $\delta {\cal H} / \delta c$ contains new terms, and 
the $\lambda$ equation thus becomes
\begin{equation}
\partial_t (-\lambda) = \nabla \cdot (\lambda \vec{v}) + \nabla \cdot 
	K \nabla c  - 
\frac{1}{2}K'(c) (\nabla c)^2 - \partial \epsilon / \partial c \ .
\label{eq:lambdaeq}
\end{equation}
Note that $\partial {\cal H} /\partial \rho $ remains the same since we have 
not included $\nabla \rho$ as noted above. From 
Eq.~(\ref{eq:vel}) one evaluates the acceleration as
\begin{equation}
\partial_t \vec{v} = \nabla \partial_t \Phi +(\lambda/\rho) 
\nabla \partial_t c + \left(\partial_t (\lambda /\rho)\right) \nabla c
\end{equation}
which yields, using Eqs.~(\ref{eq:phidot}) and (\ref{eq:lambdaeq}),
\begin{equation}
\partial_t \vec{v} + (\vec{v} \cdot \nabla ) \vec{v} =
-\rho^{-1} \nabla p + \rho^{-1} \left[ \partial \epsilon / \partial c 
-(\nabla \cdot K \nabla c) + 
\frac{1}{2} K'(c)  (\nabla c)^2  \right] \nabla c .
\label{eq:euler_c}
\end{equation}
We have  also used now the local specific enthalpy $h = h( p, c)$ and
with $\partial h /\partial p = 1/\rho.$ 
Eq.~(\ref{eq:euler_c}) means that if it is necessary to go beyond
local thermodynamics, the non-dissipative (i.e., Hamiltonian
or reactive) part of the stress tensor is modified according to
\begin{equation}
-\nabla p \rightarrow -\nabla p + \left( \frac{\partial \epsilon}{\partial c}
 -(\nabla \cdot K \nabla c) + \frac{1}{2} K'(c)  (\nabla c)^2
\right) \nabla c.
\end{equation}
This can be written in a more intuitive form. The internal energy has the form
\begin{equation}
{\cal E} = \int \left[ \epsilon(\rho, c) + g(\nabla c, c ) \right] 
d \vec{r}
\end{equation}
where the function $g$ contains derivatives of $c$. Our example above
had $g(x,y) = K(y) x^2/2.$ Now we define a generalized chemical potential,
\begin{equation}
\mu_c \equiv \delta {\cal E}/ \delta c.
\end{equation}
In terms of this chemical potential the body force is of the form 
$$
-\nabla p + \mu_c \nabla c.
$$ 
This is the general form of the
additional part of the non-dissipative part of the stress tensor
used in the body of the paper. In the critical dynamics literature 
this is known as \lq\lq Model H" in the 
Hohenberg-Halperin \cite{re:hohenberg77} lexicon.
Note that for an incompressible fluid one can rewrite this as
$$ -\nabla p' - c \nabla \mu_c $$
where $p'$ is an effective pressure chosen to guarantee 
$\nabla \cdot \vec{v} = 0$. Intuitively gradients of the chemical potential
generate body forces which need to be included in the non-dissipative
part of the stress tensor.

The equations presented here (when viscous dissipation is included)
generalize the usual Navier-Stokes
equation to situations which go beyond linear irreversible thermodynamics 
assumptions, but remain within the realm of classical fluid mechanics.

\bibliographystyle{physfla}
\bibliography{references}

\begin{thebibliography}{10}
\bibitem{re:fo1} We ignore for the time being thermal fluctuations and
possible renormalizations due to them.
\bibitem{re:fo2} In this case one may introduce an entropy functional
instead of a free energy functional along the lines described in
\cite{re:wang93} and \cite{re:alt92}.
\bibitem{re:foot_dim}
Note that one can show for Model H, by using similar arguments as in
deriving the tangential and normal surface forces in Section~\ref{sec:model}
above, that for weak temperature gradients the velocity is proportional
to both the gradient and the radius, independent of the spatial
dimensionality.
\end{thebibliography}

\end{document}